# Mid-IR fiber optic light source around 6 μm through parametric wavelength translation

A Barh[1], S Ghosh[2], R K Varshney[1], B P Pal[1, #], J Sanghera[3], L B Shaw[3] and I D Aggarwal[4]

[1]Physics Department, Indian Institute of Technology Delhi, New Delhi 110016, India
[2]Institute of Radio Physics and Electronics, University of Calcutta, Kolkata 700009, India; Currently on leave of absence at School of Physical and Mathematical Science, Nanyang Technological University, Singapore
[3]Naval Research Laboratory, Washington DC, USA
[4] Department of Physics and Optical Science, University of North Carolina at Charlotte, Charlotte NC 28223, USA

[#]Email: bppal@physics.iitd.ernet.in

**Abstract**

We report numerically designed highly nonlinear all-glass chalcogenide microstructured optical fiber (MOF) for efficient generation of light around 6 μm through degenerate four wave mixing by considering continuous wave CO laser of $5 - 10$ W power emitting at 5.6 μm as the pump. By tuning the pump wavelength, pump power, fiber dispersion and nonlinear properties, narrow (N)- and/or broad (B)- band mid-IR all-fiber light source could be realized. Parametric amplification of more than 20 dB is achievable for the N-band source at 6.46 μm with a maximum power conversion efficiency ($C_m$) $\sim 33\%$ while amplification $\sim 22 \pm 2$ dB is achievable for a B-band source over the wavelength range of $5 - 6.3$ μm with a $C_m > 40\%$.

## 1. Introduction

In recent years, a strong interest has emerged to develop fibers and fiber-based components/devices for the *eye safe* mid-IR spectral regime ($2 \sim 10$ μm) owing to potential applications in astronomy, climatology, free space communication, sensing, defense/military, medical surgery, semiconductor processing and many more [1-5]. Within this spectral range, $5 - 6.5$ μm wavelength regime is extremely utilitarian for medical diagnostics as large number of organic/inorganic molecules show their vibrational spectra in this spectral range. Compounds like As-H, HCHO, $CH_3COOH$, $CH_3$, $CCl_4$, various hydrocarbons, hydrochlorides show strong absorption in this range. Most importantly, the *absorption lines* of protein molecules like amide-I, amide-II and water molecules ($H_2O$) which are the key components of human tissues, are positioned at 6.1 and 6.45 μm. Carbon (C) presence can also be detected in this spectral regime. So, as a combine effect, this functional wavelength regime is eminently suitable for latest medical therapy like, non destructive soft/hard tissue ablation, laser surgery for brain, nerve, eye, skin etc. [5]. Therefore, it has become strategically important to develop efficient, high power, reliable light source(s) for this wavelength range.

There are indeed mid-IR sources available based on diode lasers, semiconductor quantum cascade lasers [4, 6-8], however they often require cooling system and can't handle very high power. Alternatively, all-fiber sources are more flexible to use owing to their easy portability, flexibility, high power handling capability etc. Additionally they can be coupled to optical system very easily. Two broad schemes are widely employed to realize mid-IR sources in a fiber by pumping with an available high power light source. One is via lasing action in rare earth doped fiber [1], which requires a suitable host and dopants and provide lasing at a particular wavelength. The second route involves exploiting fiber nonlinear (NL) effects [2, 9-11], which is more versatile than the first scheme as any desire wavelength could be generated with appropriate fiber design. For the fiber material, chalcogenide (S-Se-Te based) glasses are very suitable due to their high mid-IR transparency, extraordinarily linear and NL properties, low propagation loss and quite mature fabrication technologies [12-14]. On the other hand microstructured optical fibers (MOFs) with wavelength scale periodic refractive index (RI) features across the fiber cross section [15] are potentially very suitable for achieving wide tunability in dispersion and nonlinearity, which are two key characteristic parameters for nonlinearity-based new wavelength generation. Consequently, this chalcogenide glass-based MOFs (Ch-MOFs) have become quite attractive for designing fiber based mid-IR sources [2, 10, 14, 16-17].

In this paper we report our numerically designed Ch-MOF based narrow (N)- as well as broad (B)-band mid-IR sources around 6 μm by exploiting degenerate four wave mixing (D-FWM), which is the dominant NL process at proper phase matching condition [9]. Commercially available continuous wave (CW) CO laser at 5.6 μm is assumed as the input pump with $5 - 10$ W of power. Our target is to design a suitable Ch-MOF which can generate both N-band as well as B-band source in the same fiber geometry. First design was targeted to achieve an N-band source ~ 6.45 μm in a meter long fiber. Thereafter we will target to enhance the band-width (BW) for realizing B-band source in that designed fiber. Broader BW requires higher pump power ($P_0$). Accordingly, to suppress other unwanted NL effects at such high power levels, a relatively short length of fiber is targeted. Additionally, a small negative 2$^{nd}$ order group velocity dispersion (GVD) coefficient ($\beta_2$) and suitable positive 4$^{th}$ order GVD coefficient ($\beta_4$) at the pump wavelength ($\lambda_p$) would provide a broad BW at the fiber output. Finally, by suitably tuning the $\lambda_p$ and $P_0$, an efficient B-band light source extending from $5 - 6.3$ μm is theoretically realized and reported here.

## 2. Numerical modeling

Under the D-FWM process, two pump photons ($\lambda_p$) get converted to a signal photon ($\lambda_s > \lambda_p$) of higher wavelength and an idler photon ($\lambda_i < \lambda_p$) of lower wavelength when energy as well as momentum conservation is satisfied; where p, s and i, stand for pump, signal and idler, respectively. In our case the phase mismatch accumulated over linear dispersion terms ($\Delta k_L$) of these three waves are compensated by NL phase shift ($\Delta k_{NL}$) due to self phase modulation (SPM). So, the total phase mismatch ($\Delta \kappa$) becomes

$$\Delta \kappa = \Delta k_L + \Delta k_{NL} \qquad (1)$$

Now considering up to 5$^{th}$ order GVD, $\Delta \kappa$ becomes,

$$\Delta \kappa = \beta_2(\lambda_p)\Omega_s^2 + \frac{\beta_4(\lambda_p)}{12}\Omega_s^4 + 2\gamma P_0 \qquad (2)$$

where $P_0$ is pump power, $\gamma$ is the well-known effective NL parameter, $\Omega_s$ is maximum frequency shift ($\Omega_s = \omega_p - \omega_s = \omega_i - \omega_p$). These $\beta$'s are successively calculated from wavelength dependent effective RI variations. For maximum frequency shift through D-FWM process, $\Delta \kappa$ should ideally be zero. This phase matching could easily be achieved when fiber zero dispersion wavelength ($\lambda_{ZD}$) is very close to the $\lambda_p$. For negative $\beta_2$ and positive $\beta_4$ at $\lambda_p$, two pair of signal and idler will be generated. With proper amplitudes of these $\beta_2$ and $\beta_4$ at certain $\lambda_p$, it is possible to overlap the output spectrum of these two pair to realize an ultra wide BW. However, by tuning the $\lambda_p$ from this value, those pair of spectrums can be separated far from each other; hence the B-band can be discretised to realize an N-band source at furthest $\lambda_s$. The position as well as the BW of this $\lambda_s$ can also be tuned by tailoring the dispersion. Thus multi-order dispersion management is extremely crucial in this task.

For both the N-band and B-band source design, we first studied the D-FWM process under undepleted pump and lossless propagation conditions in order to get a quick estimate of the process of wavelength translation. A small idler along with the pump at the input improves the efficiency of signal generation because it induces stimulated FWM in place of spontaneous FWM to initiate the process. Under this condition, the peak amplification factor ($AF$) for the generated signal becomes.

$$AF_s = \frac{P_{s,out}}{P_{i,in}} = \left(\frac{\gamma P_0}{g}\right)^2 \sinh^2(gL) \qquad (3)$$

where $P_{s,out}$ is the output signal power, $P_{i,in}$ is the input idler power, $L$ is the fiber length and $g$ is the amplification coefficient [2, 9].

In the next step, we studied the D-FWM performance by including pump depletion and propagation loss by numerically solving the three coupled amplitude equations for variations of pump, signal and idler amplitudes ($A_j$) along z, which under CW condition are given by

$$\frac{dA_p}{dz} = -\frac{\alpha_p A_p}{2} + \frac{in_2 \omega_p}{c}\left[\left(f_{pp}|A_p|^2 + 2\sum_{k=i,s} f_{pk}|A_k|^2\right)\right.$$
$$\left. \times A_p + 2f_{ppis}A_p^* A_i A_s e^{j\Delta k_L z}\right] \qquad (4)$$

$$\frac{dA_i}{dz} = -\frac{\alpha_i A_i}{2} + \frac{in_2 \omega_i}{c}\left[\left(f_{ii}|A_i|^2 + 2\sum_{k=p,s} f_{ik}|A_k|^2\right)\right.$$
$$\left. \times A_i + f_{ispp}A_s^* A_p^2 e^{-j\Delta k_L z}\right] \qquad (5)$$

$$\frac{dA_s}{dz} = -\frac{\alpha_s A_s}{2} + \frac{in_2 \omega_s}{c}\left[\left(f_{ss}|A_s|^2 + 2\sum_{k=p,i} f_{sk}|A_k|^2\right)\right.$$
$$\left. \times A_s + f_{sipp}A_i^* A_p^2 e^{-j\Delta k_L z}\right] \qquad (6)$$

where $\alpha_j$ is the loss at the wavelength $\lambda_j$, $n_2$ is the NL index coefficient and $\Delta k_L$ is the linear phase mismatch defined in equation (1), $f_{jk}$ and $f_{ijkl}$ are overlap integrals calculated from modal field distributions [2]. Finally we have investigated the amplification factor over the entire output spectrum to find its BW.

## 3. Proposed MOF structure

All the aforementioned goals are achieved by suitably designing an As$_2$Se$_3$ based MOF structure with solid core and holey cladding that consists of 4 rings of hexagonally arranged circular holes embedded in As$_2$Se$_3$ background (cf. figure 1). Se-based glass shows good transparency in the targeted spectral range, possesses quite low material loss (~ 1dB/m) and very high nonlinearity ($n_2 \sim 10^{-17}$ m$^2$/W). Most importantly, using this Ch-glass, $\lambda_{ZD}$ can be pushed towards higher wavelengths (at the available $\lambda_p \sim 5.6$ μm). As the RI of As$_2$Se$_3$ (~ 2.79 at 6 μm) is quite high with respect to air, the dispersion slope is very large, which imposes difficulty in fine tuning of higher order GVD parameters. To tailor up to $\beta_4$ term, proper reduction in core-cladding RI difference ($\Delta n$) is needed. To fulfil it, we choose polyethersulfone (PES) to fill the holes of holey cladding. Their fabrication compatibility in various fiber forms has already been reported in the literature [18-19]. Spectral dependences of RI of both these materials [19] are incorporated in the simulations. To maintain the *effective single mode* operation, we have chosen diameter ($d$) of the hole to pitch ($\Lambda$) ratio ($d/\Lambda$) of proposed MOF < 0.45 [15]. Its single modeness is also confirmed via numerical simulation. MOF structure is also optimized in terms of minimum confinement loss ($\alpha_c$), sufficiently low modal effective area ($A_{eff}$), and proper dispersion at $\lambda_p$ suitable for generation of both N-band and B-band sources. Taking all these in consideration, we fixed the MOF parameters as $\Lambda = 2.8$ μm, radius ($r$) of hole as 0.59 μm, so that $d/\Lambda$ becomes ~ 0.42 (< 0.45). Additionally we have deliberately chosen the radius of holes in the 3$^{rd}$ cladding ring ($r_3$) to be 0.64 μm, different from the rest of the holes (cf. figure 1) to reduce $\alpha_c$ and tune dispersion.

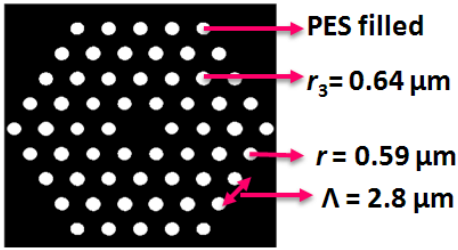

**Figure 1.** Cross section of proposed MOF. Cladding consists of 4 rings of hexagonally arranged PES filled holes (white circles) embedded in As$_2$Se$_3$ matrix (black).

For our design all the optical modes are strongly confined inside the As$_2$Se$_3$ core region, so we have only considered the wavelength dependent material loss [20] and NL parameter corresponds to this As$_2$Se$_3$ material throughout our study. $A_{eff}$ is calculated from the mode field distribution. Dispersion of the proposed MOF is plotted in figure 2, where the 2$^{nd}$ zero dispersion wavelength coincides with our operating regime. This so designed MOF structure is exploited for generation of both N-band and B-band all-fiber sources and discussed in details their characteristics in the following sections. GVD parameters as well as the modal field were calculated using the openly accessible CUDOS software in conjunction with MATLAB for numerical calculations.

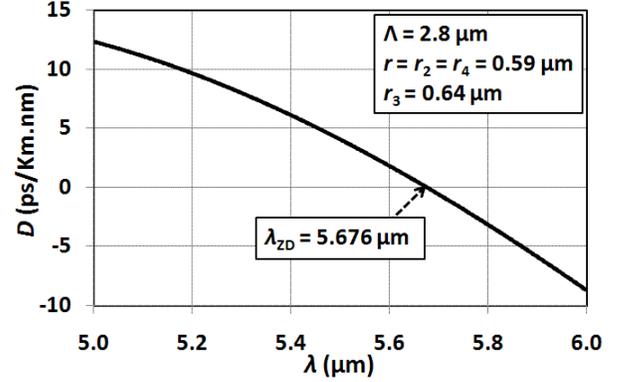

**Figure 2.** Calculated total dispersion of the designed MOF; $\lambda_{ZD} = 5.676$ μm .

## 4. N-band source generation at 6.46 μm

To get a quick estimate we first studied the D-FWM performance *without* any pump depletion condition and absence of loss by solving the coupled equations analytically. It revealed that longer fiber length ($L$) enhances the signal amplification at the expense of reduced BW. Thus there is a trade-off between these two parameters. By fixing $L$ at 1 m we have tuned the pump to study the signal output. With $P_0 = 5$ W and $\lambda_p = 5.59$ μm we achieved the best results and signal is generated at 6.46 μm (quite close to the targeted $\lambda_s$ of 6.45 μm) with maximum $AF_s \approx 20$ dB. Calculated 3 dB BW is ~ 30 nm, which makes it a relatively N-band source. The idler wave at 4.92 μm also gets amplified along with the signal. However using appropriate filter we can always filter it out depending on application.

Next we incorporated the material loss [20] and solved the three coupled amplitude equations (4)-(6) directly by considering $z$ dependent pump amplitude. For $P_0 = 5$ W, the variation of $AF$ of pump ($\lambda_p = 5.59$ μm), signal ($\lambda_s = 6.46$ μm) and idler ($\lambda_i = 4.92$ μm) waves along the propagation length are studied and shown in figure 3(a). A very small level of idler at input enhances the FWM efficiency. We have investigated the effect of this input idler power ($P_{i,in}$) by varying its amplitude from μW to mW level. It does not influence the output signal power much, however the optimum length at which power couple from pump to signal increases for lower $P_{i,in}$. For a moderate fiber length with sufficient amplification we have fixed this $P_{i,in} = 15$ mW. Figure

3(a) also indicates that the pump power transfers to signal wave at $L = 1.44$ m where the maximum $AF_s$ becomes $\sim 20$ dB. Thereafter we studied the variation of output signal power ($P_{s,out}$) around the 6.46 μm to investigate its BW (cf. figure 3(b)). The 3-dB BW of the generated signal becomes $\sim 35$ nm, which is quite close to our above-mentioned analytical calculation. It can be seen from figure 3(b), that the maximum signal power peaks to 1.64 W with power transfer efficiency ($P_{s,out}/P_0$) > 32.8%. Such a fiber if fabricated should be very attractive for certain medical diagnostics as mentioned under the introduction section.

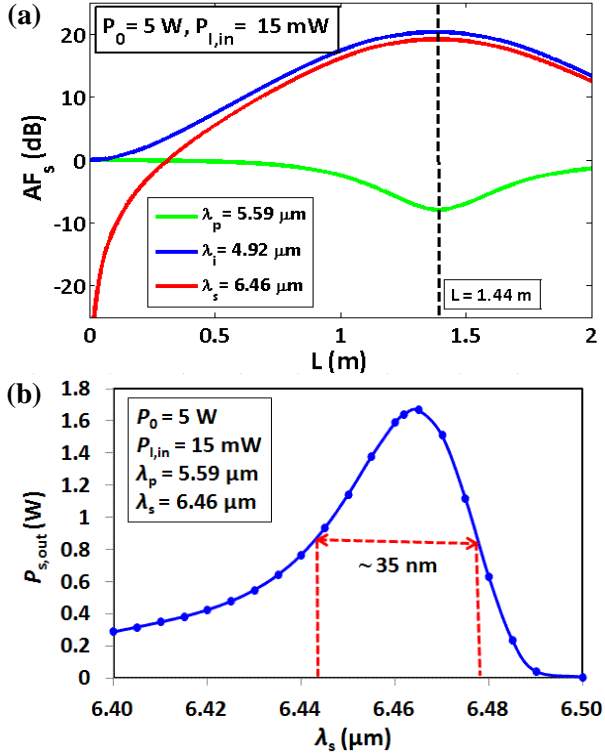

**Figure 3.** Results including pump depletion and loss; (a) variations of $AF$ for the three waves with $L$. Maximum power is coupled from pump to signal at $L \approx 1.44$ m. A small idler of 15 mW along with pump of 5 W is used at input; (b) Output power variation around the generated signal wavelength ($\sim 6.46$ μm). 3 dB BW is $\sim 35$ nm.

## 5. B-band source generation from 5 – 6.3 μm

Similar to the investigation on N-band source we first studied D-FWM performance under no pump depletion and no loss conditions. During optimization of the BW and the overall flatness of the output spectrum, a strong interplay was observed between $\lambda_p$, $P_0$, $L$, GVD parameters, and nonlinearity. Our target was to exploit the identical MOF cross section, same as N-band source, to achieve the B-band source, so that only one type of fiber could be used in a variety of applications; thus structural parameters for the MOF were kept same (cf. figure 1). As discussed earlier, dispersion management is very crucial here. Keeping these targets in mind and fixing the MOF parameters we first checked the effect of $P_0$ on output spectrum. Higher $P_0$ gives better result. We fixed $P_0$ at 10 W, however to suppress other NL effects, we need to reduce $L$ for this input power. Therefore for a lower value of $L = 50$ cm we investigated the variation of output $AF$ for different $\lambda_p$ and the same is shown in figure 4. When $\lambda_p$ coincides with $\lambda_{ZD}$, the output spectrum becomes almost flat with lower BW (black color). However, as $\lambda_p$ shifts away from $\lambda_{ZD}$, the two pairs of signal-idler spectrum start separating from each other which enhances the overlapping region and hence increases the overall BW. However further separation leads to more fluctuation in $AF$ which makes the output spectrum more discrete (the N-band source case). So, there is a trade-off between these two. We kept $\lambda_p$ at 5.61 μm for which fluctuation in $AF$ over the wavelength range of interest remains < ± 2dB and used it for further study.

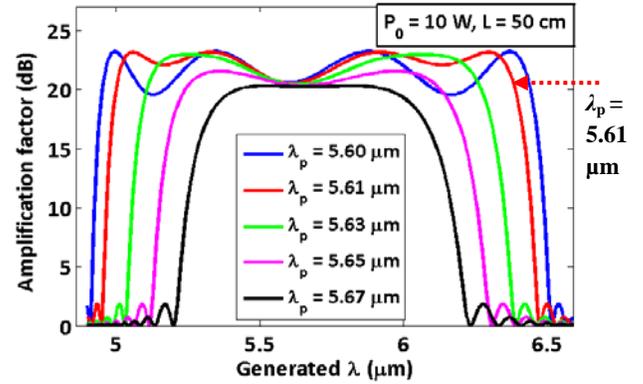

**Figure 4.** Variations of output amplification factor for different $\lambda_p$. For $\lambda_p = \lambda_{ZD}$ ($\sim 5.67$ μm), output spectrum is almost flat, centered around $\lambda_p$. As $\lambda_p$ shifts away from $\lambda_{ZD}$, the BW along with fluctuation gets enhanced.

We now incorporate the material loss [20] and solved the coupled equations directly including pump depletion, NL effects of SPM, XPM, FWM, and overlap integrals to study the D-FWM performance for B-band source generation. We have fixed $P_0 = 10$ W, $\lambda_p = 5.61$ μm and studied the variation of $AF$ for one set of three participating waves (cf. figure 5(a)). Power is coupled from pump to signal and idler at an $L = 78$ cm. Then we calculated the BW of overall output spectrum by calculating the $AF$ for it (cf. figure 5(b)). Here also the BW is almost identical with the calculated one, without pump depletion and loss (cf. figure 4). In figure 5(b) we have plotted $AF$ for 3 different input idler powers (10, 15 and 20 mW) for the corresponding coupling lengths. This plot implies that there is almost no effect of input idler power on BW and $AF$, however it effects the optimum $L$. Thus as a final result we can achieve an ultra B-band source extending from 5 $\sim$ 6.3 μm with $AF = 22\pm2$ dB and average power transfer efficiency ($P_{out}/P_0$) > 41%, covering the functional mid-IR regime, effectual for non destructive medical diagnostics and molecular spectroscopy.

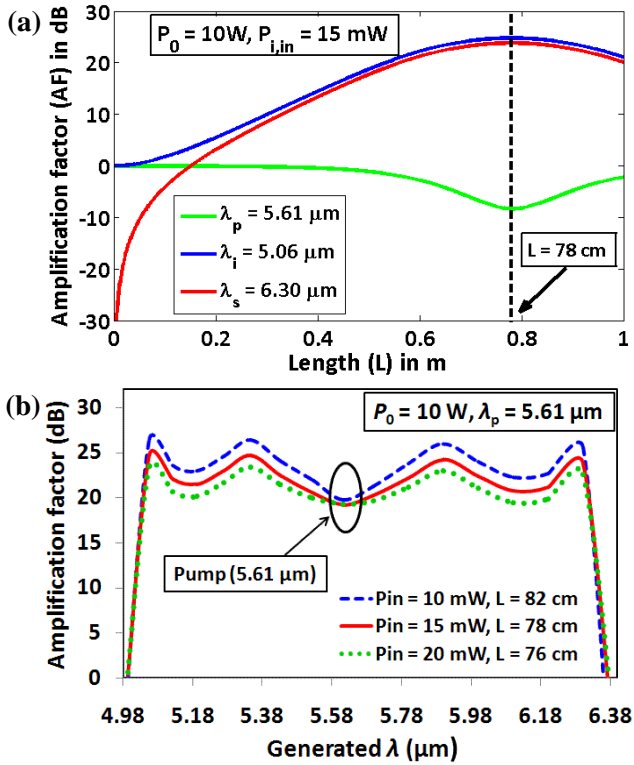

**Figure 5.** (a) Variations in *AF* for one set of pump, signal, and idler waves with *L*. A small amount of idler (15 mW) along with pump of 10 W is used at input; (b) Output spectrum for three different input idler powers along with the pump at 5.61 μm; 3 dB BW is ∼ 1.3 μm.

## 6. Fabrication challenges of the proposed MOF

To check the fabrication suitability, we investigated the tolerance of our proposed MOF structure by varying the MOF parameters ($\Lambda$, $r$, $r_3$) by few % with respect to their precisely designed values and examined sensitivity of amplification factor of the output signal ($AF_s$). For this tolerance study we outline the calculations only for the N-band source.

First we checked the effect of $\Lambda$ and $r_3$ by varying them by ± 4% and plotted corresponding variations in phase matching $\lambda_s$ and its $AF_s$ in figures 6(a)-(b), respectively. We had fixed $r$ = 0.59 μm and $L$ = 1.44 m (proposed fiber length for N-band source). Figures 6(a)-(b) imply that for ± 4% variation in $\Lambda$ and $r_3$, the position of phase matching $\lambda_s$ ($\Delta\lambda_s$) changes only by - 0.17 to + 0.12 μm and $AF_s$ remains > 19.8 dB. Thus the fabrication tolerance of these two parameters is not so critical. Then we examined the effect of variation in $r$ on the output $\lambda_s$ and its $AF_s$ and plotted the same in figures 7(a)-(b), respectively, keeping $\Lambda$, $r_3$ and $L$ unchanged. Unlike the previous one, variation in $r$ has relatively strong effect on both $\lambda_s$ and $AF_s$. From figures 7(a)-(b) it is evident that $\Delta\lambda_s$ becomes +0.34 to − 0.46 μm for only ± 1% variation in $r$ while $AF_s$ remains > 18 dB for -1 to +4% variation in $r$. So, shift in $\lambda_s$ is quite critical here.

For attaining tolerable output, variation in $r$ should be kept within ± 1%.

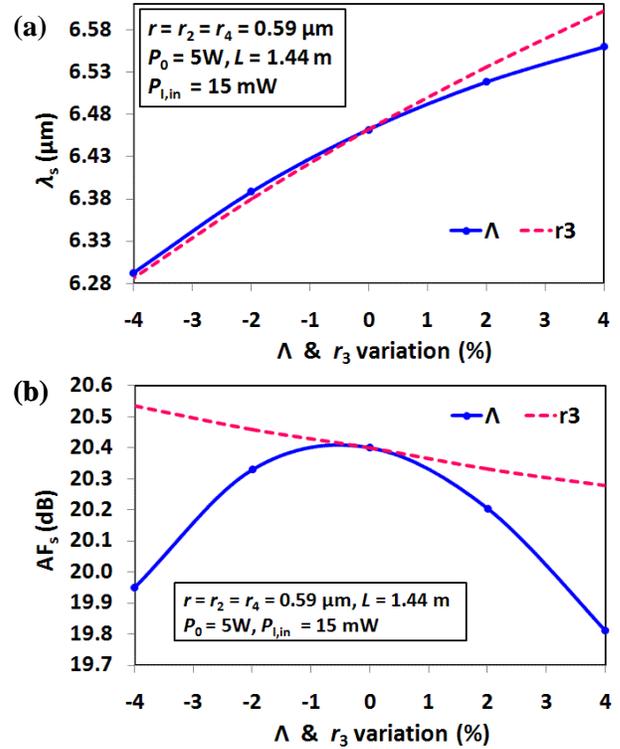

**Figure 6.** Tolerance plots by varying $\Lambda$ and $r_3$ by ± 4% for fixed $r$, $L$, $P_0$ and $P_{i,in}$. (a) variation in signal wavelength ($\lambda_s$); (b) variation in signal amplification factor ($AF_s$). Blue solid and red dashed curve correspond to the variation in $\Lambda$ and $r_3$, respectively.

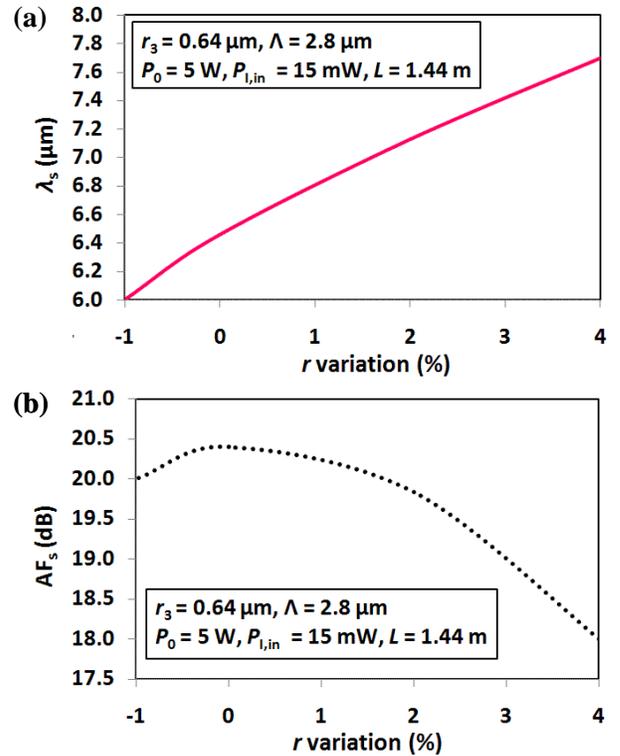

**Figure 7.** Tolerance plots by varying $r$ by -1 to +4% for fixed $\Lambda$, $r_3$, $L$, $P_0$ and $P_{i,in}$. (a) variation in signal wavelength ($\lambda_s$); (b) variation in signal amplification factor ($AF_s$).

For all the three cases as $\Lambda$, $r_3$ and $r$ increases from their proposed values, $\lambda_s$ increases (cf. figures 6(a) and 7(a)) and hence separation between three waves increases. Consequently, their modal overlap integrals ($f_{jk}$ and $f_{ijkl}$) decreases, which reduces the FWM efficiencies and hence leads to reduction in $AF_s$ (cf. figures 6(b) and 7(b)). For lower values of $\Lambda$, $r_3$ and $r$, the opposite happens, however the maximum amplification for these cases could be attained at a lower fiber length ($<$ proposed $L = 1.44$ m). Thus, $AF_s$ also decreases for lower values of these parameters at the proposed $L$. According to [21], the fabrication tolerance of MOF structural parameters can be achieved within 2–4%. Our numerical results revealed that for most of the cases the maximum limit of tolerable imperfections lie well within this limit. We can also tune the pump power and its wavelength accordingly to overcome this limit as the nature of dispersion does not change much except the position of $\lambda_{ZD}$.

## 7. Concluding remarks

In this paper, through a detailed numerical study, we have proposed a realistic design of Ch-glass based MOF structures for generation of narrow- as well as broadband watt level mid-IR fiber based light sources around 6 $\mu$m wavelength via exploitation of D-FWM process. We assumed tunable CW CO laser at 5.6 $\mu$m as the pump with 5 – 10 W of average power. With a 5.59 $\mu$m pump of 5 W at the input, N-band source at 6.46 $\mu$m with 3-dB BW of $\sim$ 35 nm can be realized in a short length (1.44 m long) fiber of our proposed MOF. The power conversion efficiency from pump to generated signal is $>$ 32.8%. On the other hand, with the same fiber cross section, an ultra B-band source ranging from 5 – 6.3 $\mu$m can be realized by tuning the pump wavelength and its power. With $\lambda_p$ at 5.61 $\mu$m and $P_0$ set at 10 W, parametric amplification as high as 22$\pm$2 dB could be achieved over the entire broad BW with a relatively high conversion efficiency $>$ 41% within an $L$ $\sim$ 80 cm of the proposed MOF. Our design optimization revolves around management of multi-order dispersion and nonlinearity for operation at the available pump sources. This wavelength range is very attractive for molecular spectroscopy of various organic/inorganic molecules and most importantly, very suitable for non destructive medical diagnostics, like soft/hard tissue ablation with minimal collateral damage.

It would be interesting to undertake the fabrication of such Ch-glass based specialty MOF for these kind of special applications where our methodology would serve as a primary design platform for new kinds of efficient mid-IR sources. Though there remain various fabrication challenges, we have shown that the tolerance of proposed MOF structure lie within the already reported fabrication limits [21]. Our proposed MOF should be realizable by so called well *matured multimaterial* fiber fabrication techniques [3, 12-14, 18].

## Acknowledgement

This work relates to Department of the Navy (USA) Grant N62909-10-1-7141 issued by Office of Naval Research Global to IIT Delhi. The United States Government has a royalty-free license throughout the world in all copyrightable material contained herein. Some salient features of these results were recently reported by us at the OSA's annual conference FiO at Orlando Fl in 2013.

AB gratefully acknowledges the award of a Senior PhD fellowship by CSIR (India). SG acknowledges financial support from the DST (India) as an INSPIRE Faculty Fellow [IFA-12; PH-13].